\documentclass[journal]{IEEEtran}

\usepackage[T1]{fontenc}
\usepackage[utf8]{inputenc}
\usepackage{amsmath, amssymb}
\usepackage{graphicx}
\usepackage{booktabs}
\usepackage{multirow}
\usepackage{cite}
\usepackage{algorithm}
\usepackage{url}
\usepackage{xcolor}
\usepackage{tabularx}
\usepackage{array}
\usepackage{longtable} 
\usepackage{array}     
\usepackage{enumitem}  
\usepackage{verbatim}
\usepackage{algorithm}
\usepackage{algpseudocode}
\usepackage{subcaption}

\title{Topological Analysis for Identifying Anomalies in Serverless  Platforms}

\author{Gianluca Reali,~\IEEEmembership{Member,~IEEE,} and Mauro Femminella,~\IEEEmembership{Member,~IEEE}
\thanks{Authors are with the Department of Engineering, University of Perugia, 06125 Perugia, Italy, and with Consorzio Nazionale Interuniversitario per le Telecomunicazioni (CNIT), 43124 Parma, Italy.}
\thanks{This work was funded by the European Union under the Italian National Recovery and Resilience Plan (NRRP) of NextGenerationEU, partnership on “Telecommunications of the Future” (PE00000001 - program ``RESTART'') and National Innovation Ecosystem (ECS00000041 - program ``VITALITY'').} 
\thanks{Corresponding author: Gianluca Reali (gianluca.reali@unipg.it).}
}

\begin{document}

\maketitle

\begin{abstract}

The information flows in serverless platforms are complex and non-conservative. This is a direct result of how independently deployed functions interact under the platform coarse-grained control mechanisms.
To manage this complexity, we introduce a topological model for serverless services. Using Hodge decomposition, we can separate observed operational flows into two distinct categories. They include components that can be corrected locally and harmonic modes that persist at any scale. Our analysis reveals that these harmonic flows emerge naturally from different types of inter-function interactions. They should be understood as structural properties of serverless systems, not as configuration errors.
Building on this insight, we present an iterative method for analyzing inter-function flows. This method helps deriving practical remediation strategies. One such strategy is the introduction of "dumping effects" to contain harmonic inefficiencies, offering an alternative to completely restructuring the service's topological model. Our experimental results confirm that this approach can uncover latent architectural structures.

\end{abstract}


\section{Introduction}

This paper investigates the operational challenges that frequently arise during the implementation and deployment of complex cloud services using serverless computing platforms \cite{faast,azure_blob}. Despite the increasing maturity of orchestration technologies and the emergence of numerous design patterns, implementing large-scale, multi-tenant cloud services in Function-as-a-Service (FaaS) mode may generate unforeseen issues that compromise correct service delivery \cite{yu2023following,saha2024survey,calavaro2025beyondcloud}.

These complications stem from multiple converging factors whose simultaneous management poses significant complexity. First, serverless architectures typically require coordinating numerous independent or loosely coupled components. Unlike traditional deployment approaches, serverless computing structures application logic as chains of elementary stateless functions, where service emerges from the distributed coordination of functions through the well-known choreography pattern \cite{Hohpe2003}. Elementary functions execute a piece of code in response to an event and remain active only transiently to serve application needs. This design minimizes network footprint, operational costs, energy consumption, and attack surface. However, these benefits come with trade-offs, such as increased code complexity, debugging difficulties, intricate function orchestration, and additional latency \cite{calavaro2025beyondcloud,10572125,Mohanty}.

A second critical factor involves interaction complexities that may generate uncontrolled loops beyond implementers' oversight \cite{awsloops}. For instance, if function is triggered by an event in a service and, during its execution, interacts with the same service, generating a new identical event and starting the cycle again, circular processes may trigger unexpectedly. Such behavior introduces delays and malfunctions whose root causes prove difficult to trace. Similarly, compensation mechanisms in design patterns \cite{Heus}\cite{orkesCompensationTransaction} introduce reverse action loops and additional invocations. For instance, in e-commerce applications, failed payments trigger compensation chains involving fund verification, card confirmation, and bank communications.
Compensation loops are an essential and designed part of Saga transactions in distributed systems \cite{GarciaMolina1987}. However, compensation loops can degenerate into implicit loops under specific pathological circumstances. They become implicit loops only when they lose their ability to terminate due to persistent errors, lack of idempotence, or bugs in the retry logic. At that point, they transform from resilience tools into cost and complexity generators.
Such loops can be detected and managed when using microservices. Instead, lack of observability in FaaS deployments, due to short function lifetime, may generate uncontrollable cycles that create logical "holes" corresponding to continuous circular data flows.

Additional complexity arises from on-demand function activation, introducing cold start latencies absent in traditional deployments \cite{10.1145/3366623.3368139,varghese,icccn25_femminella}. Despite intensive research into mitigation strategies, cold starts remain sometimes unavoidable, particularly during horizontal scaling. These events trigger cascading complications: function \textit{A} invoking function \textit{B} during cold start may encounter timeouts, leading to retry attempts that, combined with delayed \textit{B} activation, produce duplicate executions. This creates topological local curls requiring explicit management. 

Moreover, a cold start in a compensation loop can contribute to the creation of an implicit loop, or more likely, worsen an existing loop. Although it could not the primary cause of the loop, cold start can acts as a cost multiplier, making the system more vulnerable to anomalous behavior.
Further issues include backend consistency problems, cache coherence challenges, and related complications, exemplified in Table \ref{tab:topological-cycles}, which will be illustrated in detail below in the paper.

\begin{table*}[htbp]
\centering
\caption{Topological Characterization of Cycles in Distributed Serverless Architectures}
\label{tab:topological-cycles}
\renewcommand{\arraystretch}{1.2}
\scriptsize
\begin{tabularx}{\textwidth}{|
    >{\raggedright\arraybackslash}X |
    >{\raggedright\arraybackslash}X |
    >{\raggedright\arraybackslash}X |
    >{\raggedright\arraybackslash}X |
    >{\centering\arraybackslash}m{1.2cm} |
    >{\raggedright\arraybackslash}X |
    >{\centering\arraybackslash}m{1.4cm} |
    >{\raggedright\arraybackslash}X |
}
\hline
\textbf{Service / Saga Example} &
\textbf{Architecture Logical Flow} &
\textbf{Topological Model} &
\textbf{Cycle Type} &
\textbf{Betti Inv.} &
\textbf{Harmonic Component} &
\textbf{$\beta$ Reducible} &
\textbf{Observable Metrics} \\
\hline

E-commerce checkout &
Orders $\rightarrow$ Payments $\rightarrow$ Shipping $\rightarrow$ Inventory $\looparrowright$ Orders &
Nodes = functions, Edges = RPC, 2D workflow cell &
Closed transactional loop &
$\beta_0 = 1$ &
Global harmonic on $L_0$ &
No (only by breaking sync cycle) &
P99 end-to-end latency, retry rate, error amplification \\
\hline

Saga orchestration (FaaS) &
API $\rightarrow$ Validate $\rightarrow$ Enrich $\rightarrow$ Persist $\looparrowright$ Validate &
Nodes = functions, events/RPC, saga surface &
Compensation cycle &
$\beta_2 = 1$ &
Harmonic on $L_2$ &
Partial (reduce compensations) &
P99/P999 latency, compensation retries \\
\hline

ML training pipeline &
Ingest $\rightarrow$ Preprocess $\rightarrow$ Train $\rightarrow$ Validate $\looparrowright$ Preprocess &
Pipeline DAG with feedback edge &
Structural feedback loop &
$\beta_1 = 1$ &
High harmonic on $L_1$ &
No (only dampenable) &
Loop period, queue depth, training latency \\
\hline

Cold start + retries &
Invoke A $\rightarrow$ Cold start B $\rightarrow$ Timeout $\looparrowright$ Invoke A &
Functions, invocation edges &
Causal time loop &
$\beta_0 = 1$ &
Local harmonic on $L_0$ &
Yes (warm pools, backoff) &
Cold start latency, retry rate, P99 \\
\hline

Fan-out / fan-in timeout &
Orchestrator $\rightarrow$ Fan-out $\rightarrow$ Wait $\rightarrow$ Merge $\looparrowright$ Orchestrator &
Fork/join with implicit 2-cell &
Waiting surface loop &
$\beta_2 = 1$ &
Harmonic on $L_2$ &
Partial (reduce fan-out) &
Max branch latency, merge P99 \\
\hline

Eventual consistency backend &
Write $\rightarrow$ Read stale $\rightarrow$ Detect $\looparrowright$ Write &
State graph with async edges &
Convergence loop &
$\beta_1 = 1$ &
Persistent harmonic on $L_1$ &
No &
Staleness window, update lag \\
\hline

Rate-limited provider &
Request $\rightarrow$ 429 $\rightarrow$ Backoff $\looparrowright$ Request &
Endpoint with external control &
External control loop &
$\beta_1 = 1$ &
Harmonic on $L_1$ &
No &
429 rate, retry storms \\
\hline

Incoherent caches &
Read A $\rightarrow$ Update $\rightarrow$ Invalidate $\rightarrow$ Read B $\looparrowright$ Read A &
Caches as nodes, invalidation faces &
Drift surface &
$\beta_2 = 1$ &
Harmonic on $L_2$ &
Partial &
Cache hit ratio, inconsistency rate \\
\hline

Partially idempotent workflow &
F1 (idem) $\rightarrow$ F2 (non-idem) $\looparrowright$ F1 &
Nodes with side effects &
Causal duplication loop &
$\beta_0 = 1$ &
Structural harmonic on $L_0$ &
Yes (full idempotency) &
Duplicate operations \\
\hline

Autoscaling feedback &
Load $\rightarrow$ Timeout $\rightarrow$ Retry $\looparrowright$ Load &
Groups with feedback edges &
Instability loop &
$\beta_1 = 1$ &
Harmonic on $L_1$ &
Partial (predictive scaling) &
Scale oscillation, P99 spikes \\
\hline

\end{tabularx}
\end{table*}

In this context, the paper demonstrates how algebraic techniques from Topological Signal Processing (TSP) can identify operational criticalities in application execution. Leveraging Hodge theory, which establishes isomorphism between de Rham cohomology classes of differential forms and discrete Laplacian harmonic solutions on graphs, we decompose graph-defined functions into three orthogonal components: gradient, curl, and harmonic \cite{Schaub2021Tutorial}. The latter two components prove particularly relevant for analyzing application execution, as they directly associate with local service saga closures \cite{saga} and structural loops that generate inefficiencies, performance degradation, and energy consumption problems.
While Hodge decomposition provides a lens for analyzing flows, its practical efficacy is tied to the metric structure underpinning the orthogonal projection onto gradient, curl, and harmonic components. In many applications, the standard combinatorial graph Laplacian, which typically employs unit edge weights, may fail to capture heterogeneous interaction strengths or domain-specific importance, leading to a decomposition that misrepresents the true underlying flow dynamics \cite{TILLMANN2006545}. To address this limitation, in this paper we propose a procedure to weight the contributions of different edges of the network, enabling a more nuanced and accurate identification of flow components, particularly the crucial harmonic component that point out inefficiencies in function invocation chains.

The paper contributions are:
\begin{itemize}
\item Categorization of problems affecting FaaS-implemented cloud service execution
\item Development and analysis of a TSP-based model for identifying cloud service malfunction causes
\end{itemize}

Through optimized Hodge decomposition, we enable: (i) service operation diagnosis via harmonic component identification of global structural problems, (ii) automatic debugging through detection of functions participating in non-locally-eliminable load cycles, (iii) algorithm design highlighting curl and harmonic components across problem types, and (iv) definition of novel metrics, ``harmonic stress'', that increase prior to failures, remain noise-stable, and discriminate architectural variants. Hodge decomposition thus separates routable payload (gradient), local loops (curl), and structural inefficiencies (harmonic components) \cite{Schaub2020}\cite{Schaub2021Tutorial}.
The paper fundamental insight is that inefficiency causes may transcend purely architectural considerations. Two identically deployed services—with equivalent traffic, latency, and error rates—can exhibit different topological invariants requiring distinct corrective actions. Topological invariants capture architectural constraints invisible to conventional performance metrics yet fundamentally shaping system behavior. Notably, harmonic components in inter-function information flows emerge not from load itself but become visible through load, enabling targeted corrective measures.

Paper structure: Section \ref{sec:related_work} reviews existing approaches; Section \ref{sec:math_backgound} introduces algebraic and topological foundations; Section \ref{sec:proposed_method} first defines system assumptions and problem scope, then  presents our stepwise formal proposal; Section \ref{sec:experimental_results} applies our method to concrete case studies with metric quantification and discussion; Section \ref{sec:conclusion} summarizes contributions, offers concluding remarks, and outlines future directions. 

\section{Related Work} \label{sec:related_work}
This paper integrates combinatorial Hodge theory and Hodge–Laplacian techniques for analyzing data defined over graphs and cell complexes, with research on serverless and FaaS systems that represent services using dependency graphs or workflows. In the following, we outline the most pertinent strands of related work and explain how our framework—centered on a directed service graph whose nodes correspond to FaaS functions deployed in a serverless environment—relates to prior contributions in the literature.

\subsection{Combinatorial Hodge theory on graphs and edge flows}

Early mathematical groundwork for discrete Hodge theory originates in Eckmann's examination of harmonic analysis over simplicial complexes \cite{Eckmann1944}, where discrete parallels between differential forms, cohomology, and harmonic representatives were first presented. Subsequently, these concepts promoted the development of discrete exterior calculus (DEC), a systematic approach to discretizing differential operators, such as the exterior derivative, Hodge star, and Laplacian—while maintaining geometric structure \cite{Desbrun2005DEC}. Although DEC is frequently discussed in the context of meshes and manifolds, its operator-centric perspective, treating incidence matrices as discrete derivatives, aligns closely with graph-oriented Hodge decompositions applied in network analysis.

A contribution shaping contemporary approaches to ranking and inconsistency quantification on graphs is the Hodge rank framework introduced by Jiang \emph{et al.} \cite{Jiang2011HodgeRank}. In that work, pairwise comparisons are represented as an edge flow, subsequently decomposed into three orthogonal components: a \textit{gradient} part capturing globally consistent trends, a \textit{curl} part reflecting locally cyclic patterns, and a \textit{harmonic} part representing global cycles. The relative magnitudes of these components serve as indicators of different inconsistency sources. Although originally intended for statistical ranking tasks, this decomposition is applicable to any signal defined on edges, providing a conceptual basis for our use of Hodge components to disentangle interpretable function-to-function interactions from cyclic irregularities in service graphs.

Further developments have extended the methodological framework related to Hodge Laplacians and higher-order network representations. Horak and Jost \cite{HorakJost2013} investigated the spectra of combinatorial Laplace operators on simplicial complexes, linking eigenvalue behavior to topological characteristics and thereby offering theoretical justification for employing Hodge Laplacian spectra as structural descriptors. Schaub \emph{et al.} \cite{Schaub2020}, in a research contribution focused on higher-order network analysis, explored diffusion and random-walk processes governed by Hodge Laplacians, demonstrating how topological features beyond pairwise connections influence bottleneck formation and flow dynamics. These insights are pertinent to serverless contexts, where real-world deployments frequently exhibit higher-order interaction structures—such as fan-out/fan-in patterns, join nodes, and multi-function coordination—that can be represented by lifting graphs to simplicial or cell complexes.

Comprehensive overviews and tutorials have consolidated these advances into accessible analytical frameworks. Schaub \emph{et al.} \cite{Schaub2021Tutorial}, for example, offer guidance on constructing boundary operators, interpreting harmonic subspaces, and applying Hodge-based filtering to signals defined on edges and triangles. Our approach inspires to these constructions while adapting the signal definition to the specific observability metrics available in serverless environments, where edge weights may encode invocation counts, latency contributions, retry rates, or error propagation probabilities.

Extensions of signal processing methodologies from conventional graphs to higher-order topological structures have been pursued by Barbarossa and Sardellitti \cite{Barbarossa2020TSP}, who demonstrate how multi-node relationships can be examined using Hodge Laplacian operators. Subsequent work by the same authors \cite{Sardellitti} advances this line of reserch by presenting techniques for inferring cell complexes from empirical data and proposing alternative filtering strategies for flow components.

\subsection{FaaS:  performance and limitations}

The serverless computing paradigm has been extensively characterized and analyzed in foundational systems literature. Baldini \emph{et al.}~\cite{Baldini2017Serverless} define the FaaS model through its event-driven, stateless execution and granular billing mechanisms, while identifying persistent challenges including state persistence, debugging complexity, performance variability, and dependency on specific cloud providers. Hellerstein \emph{et al.}~\cite{Hellerstein2019Serverless} position serverless as a transformative evolution in cloud programming models, yet acknowledge operational issues, including constrained runtime control, performance fluctuations, and emergent system bottlenecks, that underscore the necessity for enhanced analytical and diagnostic methodologies.

Empirical investigations into platform performance have yielded substantial insights, particularly regarding cold-start phenomena, resource isolation mechanisms, and multi-tenancy effects. Wang \emph{et al.}~\cite{Wang2018Peeking} conducted systematic measurements across serverless platforms, revealing latent performance characteristics such as initialization delays and cross-provider heterogeneity. These empirical observations directly inspired our graph-oriented methodology. In particular, when a service is modeled as a directed graph of function invocations, performance degradation manifests as perturbations in edge-weighted signals and Hodge decomposition offers a mechanism for distinguishing path-consistent slowdowns from cyclical disturbances arising from retry logic or cascading failures.

Architectural innovations have accompanied these empirical studies. Akkus \emph{et al.}~\cite{Akkus2018SAND} present SAND, a serverless runtime engineered for reduced overhead through application-aware sandboxing and optimized inter-function communication. Their findings demonstrate that communication patterns among functions frequently dominate end-to-end latency, motivating the treatment of function interactions as a networked system rather than isolated components. In this case, this reinforces the suitability of Hodge-based analysis, which operates naturally on interaction graphs and can accommodate architectural modifications, such as function colocation, as adjustments to edge weights or higher-order topological structures.

Characterizations of real-world serverless workloads further substantiate the graph-centric perspective. Shahrad \emph{et al.}~\cite{Shahrad2020ServerlessInTheWild} analyze production telemetry to highlight invocation patterns, burstiness profiles, and resource consumption behaviors. Their results show the centrality of function composition and workflow topology, establishing that graph-based representations are fundamental for comprehending system-level dynamics.

The serverless deployment model has promoted research spanning theoretical developments and empirical validation. Performance-related investigations have approached the topic from multiple vantage points. Bardsley \emph{et al.}~\cite{Bardsley2018} examine evaluation methodologies, demonstrating how interconnected components induce latency across both elementary architectures and sophisticated e-commerce implementations. Nguyen \emph{et al.}~\cite{hai2019} investigate performance guarantees for real-time serverless functions, proposing a hybrid analytical-empirical model tailored to bursty, latency-sensitive workloads distributed across cloud and edge infrastructure, with validation conducted via prototype implementation.

Benchmarking methods and provider heterogeneity have received sustained attention. Copik \emph{et al.}~\cite{copik2021} introduce a comprehensive benchmarking framework that specifies workload characteristics, implementation parameters, and infrastructure configurations, yielding an abstract model of FaaS execution environments. Bortolini and Obelheiro~\cite{bortolini2020} empirically examine performance-cost trade-offs across providers, revealing pronounced dependencies on memory configuration, implementation language, and platform choice.

Platform-level innovations continue to emerge. Wang \emph{et al.}~\cite{wang2021} present LaSS, a platform designed for latency-sensitive serverless services at the edge, employing queue-theoretic resource allocation and adaptive autoscaling mechanisms incorporating fair-share policies under overload conditions. Mittal \emph{et al.}~\cite{mittal2021} propose Mu, a Kubernetes-native platform integrating core management functions—autoscaling, scheduling, load balancing—with optimized control logic; worker nodes propagate configuration and performance telemetry to a central load balancer, enabling predictive request routing and proactive workload adaptation. Jia and Witchel~\cite{jia2021} illustrate Nightcore, a function runtime distinguished by container-based isolation and microsecond-latency critical-path operations including scheduling and I/O handling.

Recent analytical advances have expanded the methodological framework. Mahmoudi and Khazaei~\cite{Mahmoudi2022, Mahmoudi2023} develop a Semi-Markov performance model focused on cold-start dynamics, validated through function invocations on AWS and KNative clusters, demonstrating accurate steady-state prediction. Panda and Sarangi~\cite{Panda2024} design an intra-node resource manager optimizing latency and fairness via L1/L2 cache monitoring and CPU wait-time analysis, incorporating reinforcement learning for dynamic priority and core allocation. Ascigil \emph{et al.}~\cite{ascigil} propose a FaaS model for edge-cloud environments, comparing centralized and decentralized resource provisioning through event-driven simulation. Pinto \emph{et al.}~\cite{pinto} formulate function placement between edge and cloud as a multi-armed bandit optimization problem, contributing a decision-theoretic perspective to resource allocation.

\subsection{Bridging Hodge decomposition with FaaS observability}

To the extent of our knowledge, the unique explicit applications of combinatorial Hodge decomposition to serverless service graphs is \cite{hodge_submitted}. However, that paper does not explore any form of algebraic adaptation for the correct isolation of critical flow components, as proposed in this paper. Nevertheless, a strong conceptual alignment exists. HodgeRank-style decompositions \cite{Jiang2011HodgeRank} were originally formulated to detect inconsistency and cyclic structure in edge-valued data, and serverless telemetry naturally yields edge-weighted measurements—such as invocation counts, latencies, or error rates—defined over function-call graphs. Furthermore, higher-order generalizations employing simplicial complexes \cite{Schaub2020, Schaub2021Tutorial} enable the representation of multi-function coordination patterns which are challenging to capture through pairwise edges alone. 

In this context, our paper advances the state of the art by: (i) establishing a systematic mapping from serverless observability data to signals defined on edges and higher-order faces; (ii) applying Hodge decomposition to disentangle globally consistent trends from locally and globally cyclic behaviors; and (iii) redistributing the harmonic component toward the edges affected by the observed performance issue so as to make the observation a reliable picture of the energy trapped in unattended cycles. In doing so, we position Hodge decomposition as a lightweight and interpretable enhancement to serverless observability, offering a complementary perspective to black-box anomaly detection techniques and workflow-level optimization strategies.

\section{Background and Topological Model}  \label{sec:math_backgound}

Consider an oriented graph graph $G=(V,E)$, where $V = \{v_0, \dots, v_{N-1}\}$ is the set of vertices, or nodes, and $E = \{e_0, \dots, e_{M-1}\}$ is the set of oriented edges connecting node pairs. In our model each vertex is associated with a deployed and running serverless function, and each edge is associated with function invocation from the start node to the end node connected by the oriented edge. Thus it can be regarded as the flow of information exchanged between the connected functions. A circular set of vertices can be associated with a managed circular invocations of function over the loop. Such a loop is also referred to as a saga, and is represented over the graph as a $face$. If an operation fails, the saga behavior includes compensatory actions to manage the failed operation.

Formally, this service invocation graph $G$ is associated with a \textit{cellular complex} $\mathcal{K}$. It is a finite collection of cells of dimension $k=0,1,2$. Each $0$-cell corresponds to a vertex in $V$; Each $1$-cell is an edge $e$ associated with an ordered pair of vertices, i.e. a logical relation between two $0$-cells $(v_i,v_j)$. A $2$-cell is associtaed with a face, and is attached along a closed chain of $1$-cells with a coherent orientation, (e.g. clockwise). 
We use cellular complexes instead of the more popular simplicial complexes since not every subset of functions and invocations can be used to define a saga.

On this topological structure, the spaces of the co-chains are defined as follows.
A k-cochain set $C^k$ includes the real-valued functions defined on k-cells. Specifically:

\begin{itemize}
\item $C^0 (\mathcal{K})=\{\alpha:V\xrightarrow{}\mathbb{R}\}$: 0-cochain functions defined over nodes.  

\item $C^1(\mathcal{K})=\{\omega:E\xrightarrow{}\mathbb{R}\}$: 1-cochain functions defined over edges, with $\omega(-e)=-\omega(e)$)

\item $C^2(\mathcal{K})=\{\sigma:F\xrightarrow{}\mathbb{R}\}$:  2-cochain functions defined over faces.  
\end{itemize}

\textit{Differential operators} on a graph, also referred to as \textit{co-border operators}, are used for discrete analysis to transpose the concepts of continuous calculus (such as gradient, divergence and Laplacian) to discrete structures. They allowing analyzing variations, flows and connectivity between the elements of the cellular complex defined above.
They are formally defined as linear maps that increase the degree of the chain over which the co-chain is defined. For example the operator
$d_0 : C^0 \to C^1$ returns the discrete gradient of the graph. In particular, applying the operator to a function $\alpha$ defined on the vertices of an oriented edge $e=u \to v$, it returns $(d_0 \alpha)(e) = \alpha(v) - \alpha(u)$, which corresponds to the measure of the potential difference along the edge. Similarly, the differential operator $d_1 : C^1 \to C^2$ is an operation of a discrete rotor on the graph. In fact, considering an oriented face $\phi$ (e.g. counterclockwise) with edges $e_1,e_2,…,e_m$ coherently oriented, it returns:
\begin{equation}
(d_1 \omega)(\phi) = \sum_{i=1}^{m} \omega(e_i)
\end{equation}
The edge operators over graph take the form of incidence matrices:
\cite{Schaub2021Tutorial} and \cite{Barbarossa2020TSP}:

\begin{itemize}
    \item $B_1 \in \mathbb{R}^{N \times M}$ (nodes $\to$ edges)
    \item $B_2 \in \mathbb{R}^{M \times F}$ (edges $\to$ faces or cells)
\end{itemize}

For the construction of incidence matrices, given the cellular complex $\mathcal{K}$ it suffices to enumerate functions (index set $K_0$), enumerate directed service interactions (index set $K_1$) and enumerate higher-order interaction motifs (index set $K_2$).

\begin{equation}
(B_1)_{v,e} =
\begin{cases} 
+1 & \text{if } v \text{ is the head of } e \\
-1 & \text{if } v \text{ is the tail of } e \\
0 & \text{otherwise}
\end{cases}
\end{equation}

$B_2$ is defined analogously using consistency between edge orientation and direction of passage of the edge in the circulation of the face. Both matrices are clearly sparse. By their definition, it follows that $B_1 B_2 = 0$. Since $B_1$ encodes the relationship between functions and calls, and $B_2$ encodes the relationship between calls and saga workflows, $B_1 x$ can be regarded as the divergence of any call flow, while $B_1^T x$ represents the flow circulation, e.g., loops and cyclic retries.

In this paper we do not consider the any differential operator $d_2 : C^2 \to C^3$, which could model the divergence on 3D volumes.

In the canonical space $ img (\mathbf{1})$ the \emph{scalar product} over the set $C^k$ takes the usual form:
\begin{equation}
\langle \alpha, \beta \rangle_{C^k}
=
\sum_{c \in \{k\text{-cells}\}} \alpha(c)\,\beta(c).
\end{equation}

The \emph{adjoint operator} of $d_k$ is $d_k^* : C^{k+1} \to C^k$ such that
$\langle d_k \alpha, \beta \rangle_{C^{k+1}}$ =
$\langle \alpha, d_k^* \beta \rangle_{C^k},
\forall \alpha \in C^k,\ \beta \in C^{k+1}.$
In particular,
$d_0^* : C^1 \to C^0$  is associated with the discrete divergence on the nodes of the
graph. In particular, given a vertex $v$, it is immediate to calculate
\begin{equation}
(d_0^* \omega)(v)
=
-
\sum_{\substack{\text{edges } e \\ \text{entering } v}} \omega(e)
+
\sum_{\substack{\text{edges } e \\ \text{exiting } v}} \omega(e).
\end{equation}

\noindent
or with a different sign depending on convention; often $d_0^* = -\mathrm{div}$. Using these definitions, the Hodge Laplacian is defined as
$L_k (C^k \to C^k)= d_{k-1} d_{k-1}^* + d_k^* d_k$.  

By using the matrix definition of differential operators, the Hodge laplacian for $K=0,1,2$ is:
\begin{equation}
L_0 = B_1 B_1^T
\end{equation}
\begin{equation}
L_1 = B_1^T B_1 + B_2 B_2^T
\end{equation}
\begin{equation}
L_2 = B_2^T B_2
\end{equation}

We consider $L_1$ since flows generated by function calls are modeled by 1-cochains. Let it therefore be $f \in C^1(\mathcal{K})$ an observed flow (e.g., calls, error rate). The central result of the Hodge theory is the Hodge decomposition \cite{Schaub2021Tutorial}. According to it, every flow $f\in C^1$ admits an orthogonal decomposition:

\begin{equation}
f = B_1^T \phi + B_2 \psi + h 
\end{equation}

Being $\phi$ and $\psi$ potentials defined on nodes and edges which generate the irrotational and solenoidal components of the flow $f$. They can be found by projecting $f$ over $Img(B_1^T)$ and $Img(B_2^T)$, respectively.  These three components are:
\begin{itemize}
    \item \textit{Gradient component}  $B_1^T \phi$: it is due to flows generated by “potential” differences between nodes (e.g. a function calls another because the latter offers a resource, similar to a pressure gradient).
\item  \textit{Rotational} or \textit{curl component} $ B_2 \psi$: it reveals call loops between functions due to feedback or designed loops. It is useful to highlight circular dependencies between functions in sagas.
\item \textit{Harmonic component h}: it is divergence-free, $B_1h=0$, and curl-free, $B_2^Th=0$. Hence, this component has the fundamental property that $L_kh = 0$. 
\end{itemize}
The importance of the harmonic component comes from the isomorphism, shown by the Discrete Hodge Theorem \cite{Dodziuk1976},  between kernel $\mathcal{H}^k = \ker(L_k)$ and the $k$-cohomology, defined as
\begin{equation}
H^k
=
\frac{\ker \bigl(d_k : C^k \to C^{k+1}\bigr)}
     {\operatorname{im} \bigl(d_{k-1} : C^{k-1} \to C^k\bigr)}.
\end{equation}

$H^k$ is the space of closed $k$-forms modulo the exact ones.
More specifically, the $2$-cohomology groups includes independent cycles identifiable on the graph, which differ at most
by a gradient (i.e., a cycle that is the edge of a face).
Therefore, the fundamental results is that the $2$-th group of cohomology includes paths in the graph that represent ``holes'', i.e.\ those cycles that cause problems and inefficiencies. These holes, representing unmanaged flow of calls potentially generating operational problems, can be found in the kernel of the Laplacian. Moreover, given a class \([\omega] \in H^k\) (where it is closed, \(\omega: d_k \omega = 0\)), there is only one \(h \in [\omega]\) such that:
\begin{equation}  
d_1 h = 0 \quad (B_2^Th=0) \quad \text{and} \quad d_{0}^* h = 0 \quad (B_1h=0)
\end{equation}

Furthermore, it is worth to mention that the kernel size of Hodge Laplacian corresponds to the Betti Numbers, which are related to fundamental invariants \cite{Friedman1998} (examples in Table \ref{tab:topological-cycles}):
\begin{itemize}
    \item $\beta_0 = \dim \ker L_0$, is related to connected components of the graph. In other words, it indicates the number of independent subgraphs. In serverless platforms, it can indicate the isolated functions, or groups of them, that do not communicate.
    \item $\beta_1 = \dim \ker L_1$, is related to structural cycles. A nonzero value of $\beta_1$ indicates the number non-contractable, global loops. In the serverless case, it can reveal cyclic dependencies, which may lead to structural fragility.
    \item $\beta_2 = \dim \ker L_2$, is related to closed workflows. It measures the number of nontrivial two-dimensional cohomology classes. 
In a serverless platform, a nonzero $\beta_2$  reveals globally closed execution structures, such as multi-service transactions or compensation workflows.
\end{itemize}

Beyond the mere presence of a harmonic component, its energy $\| h \|^2$ is also important since it provides a quantitative measure of the global harmonic load affecting the service architecture. 

Some insights can also be obtained from the spectrum of the Laplacian. Its eigenvalues of \( L_k \) are unaffected by local disturbances and change only with structural changes. Typically, small gaps indicate a fragile system, and large gaps indicate robust one.

\section{Proposed Method} \label{sec:proposed_method}
\subsection{Research Objective} \label{sec:problem_definition}
\textit{The research objective is to identify architectural weaknesses in the services deployed on serverless platforms.} 

Many architectural weaknesses that can affect a Serverless/FaaS system. After proposing a general method, due to space limitations, we focus on analyzing how cold starts interact with the possible fragility of a compensation system. We assume the presence of bugs or a temporarily unstable dependency that generates failures. For example, if the execution environment of the involved functions is prone to failures, timeouts may occur and the serverless platform must perform a cold start, creating a new environment, loading code and dependencies. Cold starts introduce additional delay into an operation that is already in a critical phase and this delay can cause further delays and timeouts. If the retry logic is not idempotent, it can generate duplicates, and activate retry mechanisms at higher levels (e.g., the message queue that retries the sending), creating a chain reaction. In practice, cold start adds \textit{inertia}  and \textit{latency} that could generate self-perpetuating loops that fail to terminate. With reference to the Hodge decomposition, harmonic components are generated to the function invocation flow. 

Such a serverless architecture can be modeled as a cellular complex $\mathcal{K}$, a sequence of observed operational flows $\{f(t)\}_{t=1}^T \subset C^1(\mathcal{K})$. Such flows can be decomposed into gradient, curl, and harmonic components, and structural issues and inefficiencies be associated with them.

Although this information is already useful in itself, its direct use to correct services in operation requires further insights. In particular, the following aspects should be considered.
\begin{itemize}
    \item Cochains on the invocation graph are not conservative flows, since functions could terminate services or generate new calls downstream.
    \item Flows defined through operational quantities collected metric servers, such as requests per second (RPS), CPU usage, memory usage, and latency, are partial and time-limited picture of the operation of the service. 
    \item Measurement errors and stochastic behavior may happen.
\end{itemize}
Therefore, it could happen that the result of the Hodge decomposition is affected by spurious harmonic components that appear on some edges, i.e. harmonic residues of numerical origin that could be explained by other components, i.e. gradient or curl. If this happens, it means that the problem on those edges is not intrinsically topological but is an artifact of the discrimination capacity of the metric space used.

For this reason, \textit{the specific research objective is to find and introduce a realistic metric, through which the harmonic component vanishes on the edges that are not part of routes that are structurally problematic}. In other words, the information flow that passes through them is explainable and manageable. The residual harmonic component instead focuses on loops which represent real points of architectural fragility.

The peculiarity of Hodge theory lies in its decomposition of a flow into orthogonal components, a separation that is fundamentally dependent on the choice of an inner product—the Riemannian metric on a manifold or its discrete analogue, a weight matrix on a graph. In discrete settings, the default choice often assigns uniform weight to every edge, an assumption that is rarely valid in complex systems like serverless architectures or traffic networks where interactions possess vastly different strengths, latencies, or costs. A uniform metric can obscure the relative importance of specific functional dependencies or data pathways, potentially misclassifying a heavily weighted cyclic interaction as a negligible curl if its magnitude is small, or conversely, overstating the significance of low-importance but high-magnitude flows. For this reason, in what follows we develop a procedure to identify a suitable metric to highlight the actual inefficiencies of the function invocation chains.

\subsection{Hodge decomposition}

We assume that the service topology is fixed over the observation window, flows are aggregated over time intervals sufficient to ensure stationarity. Function interactions are inferred from execution traces. The observable quantities, such as latency, number of calls, and errors, can be used to define flows on the invocation graph. Relaxing these assumptions and extending the framework to dynamic topologies is left for future work.

The harmonic component of the Hodge decomposition can capture the structural inefficiencies in serverless architectures that cannot be mitigated through local load balancing or retry control mechanisms.
Although no flow conservation can be assumed, since an incoming request to a function can be either processed and and routed downstream, or  terminated, or amplified by generation of a number of output requests, we can anyway decompose this dynamic flow as follows.

First of all, $B_1, B_2$ matrices derive from the implemented functions calls and APIs. 

The gradient component can be determined as follows.
In a FaaS system, the net incoming flow is $> 0$ due to incoming requests. Also a net outgoing flow $> 0$ is present due to sinks. Thus we can identify components  that not belong to $\operatorname{im}(B_1^T)$ (the space of conservative flows). We want to find $\phi$ such that the gradient component $f_{\text{grad}}$ is the orthogonal projection of $f$ on the space $\operatorname{im}(B_1^T)$. The gradient component lies in $\operatorname{im}(B_1^T)$ and is obtained by solving:
\begin{equation}
\min_\phi \| f - B_1^T \phi \|^2 = f^\top f - 2f^\top B_1^T \phi + \phi^\top B_1 B_1^\top \phi,
\end{equation}
Thus, deriving with respect to $\phi$, $\nabla_\phi J(\phi) = -2B_1 f + 2B_1 B_1^\top \phi = 0$ the normal equations $B_1 B_1^\top \phi = B_1 f$ are obtained, since $L_0 \phi = B_1 f$ and $L_0 = B_1 B_1^\top$. Since $L_0$ is singular and $L_0$ always has the zero eigenvalue with eigenvector $\mathbf{1} = (1, 1, ..., 1)^\top$, the system $L_0 \phi = B_1 f$ has a solution only if $B_1 f$ is orthogonal to $\mathbf{1}$, i.e. $\mathbf{1}^\top (B_1 f) = 0$. But since $\mathbf{1}^\top B_1 = 0$ by construction, the system has infinite solutions.
To make the solution unique, we used 
the Moore–Penrose pseudoinverse $L_0^+$:
    \begin{equation}
        \phi = L_0^+(B_1 f)
    \end{equation}
    which provides the minimum norm $\| \phi \|$ solution and satisfies $1^\top \phi = 0$ if $B_1 f$ is in the range of $L_0$. 
Thus, the gradient flow is $f_{\mathrm{grad}} = B_1^\top \phi$.

The curl component lies in $\mathrm{im}(B_2)$ and is obtained by solving:
\begin{equation}
    \min_{\psi} \| f - B_2 \psi \|^2
\end{equation}
The normal equations give $B_2^\top B_2 \psi = B_2^\top f$, that is $L_2 \psi = B_2^\top f$, with $L_2 = B_2^\top B_2$.
Then, $f_{\mathrm{curl}} = B_2 \psi$.

Finally, the harmonic component is  $h = f - f_{\mathrm{grad}} - f_{\mathrm{curl}}$, which is also the orthogonal projection of $f$ onto $\ker(L_1)$.  

Determining the Hodge components for services implemented by using some tens of functions does not cause any particular scalability issues. If the analysis is extended to the datacenter level, including thousands of functions, to solve the most complex step $L_0 \phi = B_1 f$ it could be necessary to use iterative methods \cite{SpielmanTeng2014}.

\subsection{Metric identification}\label{sec:metric}

It is worth to note that some metrics can be easily mapped on nodes, i.e.\ associated with functions (e.g., CPU or memory usage by function, cold start frequency). Some other metrics are naturally mapped on edges, i.e. the calls (e.g. invocations between functions, network latency, error rate of communication). 
For edge-defined metrics, the decomposition can be immediately applied. For node-defined metrics, co-chains are defined on nodes. In this case, instead of applying Hodge decomposition to raw metrics, it is necessary to use derived quantities computed by suitable mappings for flow definition over edges. In fact, if we simply define $f = B_1^\top x \Rightarrow f \in \mathrm{im}(B_1^\top)$ then for Hodge decomposition 
$f_{\mathrm{grad}} = f$ and $f_{\mathrm{curl}} = 0$, $f_{\mathrm{harm}} = 0$. Therefore, it is necessary to define $f \in \mathbb{R}^M$ as a mutual influence between the functions based on the type of desired analysis, then Hodge decomposition to $f$ can be applied. 

Given the classical analysis based the Hodge decomposition, the potential issues mentioned in Section \ref{sec:problem_definition} can be faced through a topological approach. The topological approach, much more radical, consists in a change in the topology of the network to alter the homology classes and the relative Betti numbers of the graph. This requires reviewing the logic that oversees the function calls. To facilitate this type of solution, we can combine the topological approach with a geometric approach, shown in this section. It is certainly more immediate, and although it does not alter the topology of the graph, with the respective homology classes, it allows obtaining useful and usable information with few processing steps. In practice, while leaving the topology unchanged, and therefore all the components deriving from the Hodge decomposition, it is possible to operate geometrically to transfer the energy content between the various Hodge components though a base modification. The purpose is to deepen the edge analysis to identify whether the harmonic components of the flow persists in the new base or can be explained coherently as part of a saga (curl) or as a gradient of a potential, once a metric that adequately illustrates the real operation is used.
To do this, we adapt the scalar product operation, only from the point of view of magnitude, while preserving the orthogonality of the Hodge components. To do this, it is sufficient to carry out a coordinate transformation by simply changing the units of measurement for each dimension considered.

Consider again the running example. The co-chains deriving from the Hodge decomposition belongs to \( C^1(\mathcal{K})\), i.e. they are defined over the edges of the graph.

Let \( E \), the set of edges, be a normal base of \( \mathbb{R}^M \), with column vectors \( e_0, \dots, e_{M-1} \), and let \( P \) the invertible \( m \times m \) matrix with columns \( e_0, \dots, e_{M-1} \). 
If \( x \) are the coordinates in the canonical base and \( x' \) are the coordinates in the new base \( E \), then \( x = Px' \) e \( x' = P^{-1}x \). Suppose we want the scalar product \( \langle \cdot, \cdot \rangle_{M_1} \) become the standard scalar product \( \langle u, v \rangle_{M_1} = (u')^Tv' \) in the new base, being $M_1$ the metric matrix of the scalar product. We look for a base \( E \) (matrix \( P \)) such that:  
$ \langle u, v \rangle_{M_1} = (u')^T v'$ if $u = Pu', v = Pv'$ 
where $u', v'$  are the coordinates in that base. Therefore:
$ \langle u, v \rangle_{M_1} = u^TM_1v = (Pu')^T M_1(Pv') = (u')^T (P^T M_1 P)v' $.
We want it to be \((u')^T v'\), so we must have \(P^T M_1 P = I\), i.e. \(M_1 = (P^{-1})^T P^{-1}\). The natural choice is to take \(P = M_1^{-1/2}\), with \(M_1^{1/2}\) is the positive definite symmetric square root of \(M_1\). Clearly \(M_1^{-1/2}\) is symmetrical if \(M_1^{1/2}\) is too. So the change of coordinates that reformulates the scalar product is:
\(y = M_1^{1/2} x\).

Therefore, we can introduce a mass matrix on \(C^1\), of size \(|E| \times |E|\), symmetric and positive definite. If we choose \(M_1 = \text{diag}(m_e)\), where \(m_e > 0\) is a weight for each edge, we preserve the orthogonality of the base vectors. In other words, what happens over an edge is not coupled with neighboring edges.
\textit{Node criticality is encoded through an edge-based metric \(M_1\), inducing a weighted Hodge decomposition where flows traversing critical functions contribute more to the system energy.} 

Given the general expression of the Hodge Laplacian
\(L_k = d_{k-1} d_{k-1}^* + d_k^* d_k\)
in the reformulated scalar product, we want to find the adjunct \(d_0^\dagger\) such that
\(\langle d_0 \phi, f \rangle_1 = \langle \phi, d_0^\dagger f \rangle_0\). For what concerns the left hand side,
\(\langle B_1^T \phi, f \rangle_{M_1} = (B_1^T \phi)^T M_1 f = \phi^T B_1 M_1 f\). Considering the right hand side, \(\langle \phi, d_0^\dagger f \rangle_0 = \phi^T (d_0^\dagger f)\). Thus,
\(\phi^T B_1 M_1 f = \phi^T (d_0^\dagger f) \forall \phi\). 
Hence \( d_0^\dagger = B_1 M_1 \).

We consider now \( d_1 = B_2^T \). We want to find \( d_1^\dagger \) such that
\begin{equation}
   \langle d_1 f, \omega \rangle_2 = \langle f, d_1^\dagger \omega \rangle_1
\end{equation}
For the heft hand side \(\langle B_2^T f, \omega \rangle_2 = (B_2^T f)^T M_2 \omega = f^T B_2 M_2 \omega\), where $M_2$ is a mass matrix for scalar product on 2-forms. For the right hand side,
\(\langle f, d_1^{\dagger} \omega \rangle_1 = f^T M_1 (d_1^{\dagger} \omega)\), thus
\begin{equation}
    f^T B_2 M_2 \omega = f^T M_1 (d_1^{\dagger} \omega) \quad \forall f
\end{equation}
Hence 
\begin{equation}
M_1 (d_1^{\dagger} \omega) = B_2 M_2 \omega.
\end{equation}

Multiplying by \( M_1^{-1} : d_1^{\dagger} \omega = M_1^{-1} B_2 M_2 \omega \) i.e.: \( d_1^{\dagger} = M_1^{-1} B_2 M_2 \).

Using the reformulated scalar product the Hodge Laplacians with \( M_1 \) thus become:
\begin{align}
L_0 &= B_1 M_1 B_1^T \\
L_1 &= B_1^T M_1 B_1 + M_1^{-1} B_2 M_2 B_2^T \\
L_2 &= B_2^T M_1 B_2
\end{align}

The gradient projection we want to find is
\begin{equation}
\min_{\phi} \| f - B_1^T \phi \|_{M_1}^2
\end{equation}
\noindent
which leads to:
\begin{equation}
B_1 M_1^{-1} B_1^T \phi = B_1 M_1^{-1} f
\end{equation}

The curl projection requires finding
\begin{equation}
    \min_{\psi} \| f - B_2 \psi \|_{M_1}^2,
\end{equation}
\noindent which gives:
\begin{equation}
B_2^T M_1^{-1} B_2 \psi = B_2^T M_1^{-1} f
\end{equation}
The harmonic component is obtained by subtracting the two previous ones from the total flow.

\subsection{Optimal detection of harmonic components}
The open aspect in the model illustrated in the previous section is the determination of $M_1$, which better allows identifying the harmonic components to be controlled.
For this purpose, consider a reference metric $M_{ref}=diag(m_e^{ref})$, $m_e^{ref} > 0,  \forall e \in E$. This metric represent a baseline against which deviations are measured. By using this metric we define the cost functional
\begin{equation}
 J(M_1)=\|h(M_1)\|_{M_1}^2+\lambda tr(M_1)+\beta\|M_1-M_{ref}\|^2
\end{equation}
\noindent
with $\|h(M_1)\|_{M_1}^2=h^TM_1h$ 
and $\|M_1-M_{ref}\|^2=\sum_e(m_e-m_e^{ref})^2$, with the boundary   $m_e > m_{min}>0 \quad \forall e \in E$,
whereas $h(M_1)$ is the harmonic component of the Hodge decomposition calculated with the metric 
$M_1$.

Since \textit{h} depends on $M_1$, the Karush–Kuhn–Tucker (KKT) conditions of the complete problem would be very complex. The following consistent recursive strategy is then used:

\begin{itemize}
    \item fix $M_1^{(k)}$ and compute $h(M_1^{(k)})$;
    \item fix  $h(M_1)^{(k)}$ and minimize $J(M_1^{(k+1)})$ over $M_1^{(k+1)}$, according to the procedure below and Algorithm \ref{alg:complete-recursive-iteration}.
\end{itemize}

The first subproblem in $M_1$ can be written:
\begin{equation}
    \min_{m_e \geq m_{\text{min}}} \sum_{e\in E} \left[ m_e \left( h_e^{(k)} \right)^2 + \lambda m_e + \beta \left( m_e - m_e^{\text{ref}} \right)^2 \right]
\end{equation}

This problem formulation is clearly convex and separable for each edge. The lagrangian can be written as:
\begin{align}
\mathcal{L}(m, \mu) = & \sum_e \left[  m_e \left( h_e^{(k)} \right)^2 + \lambda m_e + \right. \\ \nonumber 
& \left. +\beta \left( m_e - m_e^{\text{ref}} \right)^2 - 
\mu_e \left( m_e - m_{\text{min}} \right) \right]
\end{align}

The KKT conditions, computed for each edge, give:

\begin{equation}
m_e \geq  m_{min}
\end{equation}
\begin{equation}
\mu_e \geq  0
\end{equation}
\begin{equation}
\label{eq:stationarity}
\frac{\partial \mathcal{L}}{\partial m_e} = \left( h_e^{(k)} \right)^2 + \lambda + 2\beta\left( m_e - m_e^{\text{ref}} \right) - \mu_e = 0
\end{equation}
\begin{equation}
\mu_e (m_e-m_{min})=0
\end{equation}

From (\ref{eq:stationarity}) we obtain
\begin{equation}
m_e = m_e^{\text{ref}} - \frac{\left(h_e^{(k)}\right)^2 + \lambda - \mu_e}{2\beta}
\end{equation}.

Two cases are distinguished. 
In case of inactive constraint:
\begin{equation}
m_e > m_{\text{min}} \Rightarrow \mu_e = 0
\end{equation}
\begin{equation}
m_e^* = m_e^{\text{ref}} - \frac{\left(h_e^{(k)}\right)^2 + \lambda}{2\beta}
\end{equation}

In case of active constraint:
\begin{equation}
m_e = m_{\text{min}} \Rightarrow \mu_e \geq 0
\end{equation}

The overall solution can be written as:
\begin{equation}
m_e^{(k+1)} = \max  \left( m_{\text{min}}, m_e^{\text{ref}} - \frac{\left(h_e^{(k)}\right)^2 + \lambda}{2\beta} \right) 
\end{equation}

\begin{algorithm}[t]
\caption{Complete recursive iteration}
\label{alg:complete-recursive-iteration}
\begin{algorithmic}[1]
\Require Reference metric $M_{\mathrm{ref}}$, lower bound $m_{\min}$,
         parameters $\beta>0$, $\lambda \ge 0$, tolerance $\varepsilon>0$
\Ensure Sequence of metrics $\{M_1^{(k)}\}_{k\ge 0}$

\Statex \textbf{Initialization}
\State $M_1^{(0)} \gets M_{\mathrm{ref}}$

\For{$k = 0,1,2,\dots$}

  \Statex \textbf{Hodge decomposition with metric $M_1^{(k)}$}
  \State $f \gets f_{\mathrm{grad}}^{(k)} + f_{\mathrm{curl}}^{(k)} + h^{(k)}$

  \Statex \textbf{Metric update}
  \State $m_{e}^{(k+1)} \gets 
    \max\!\left(
      m_{\min},
      m_{e}^{\mathrm{ref}} -
      \dfrac{\bigl(h_{e}^{(k)}\bigr)^2 + \lambda}{2\beta}
    \right)$
  \State $M_1^{(k+1)} \gets \mathrm{diag}\!\bigl(m_e^{(k+1)}\bigr)$

  \Statex \textbf{Stopping criterion}
  \If{$\dfrac{\lVert m_e^{(k+1)} - m_e^{(k)} \rVert}{\lVert m_e^{(k)} \rVert} < \varepsilon$}
    \State \textbf{break}
  \EndIf

\EndFor
\end{algorithmic}
\end{algorithm}

As shown in Algorithm~\ref{alg:complete-recursive-iteration}, 
the metric is iteratively updated using the harmonic component
of the Hodge decomposition.

\section{Experimental Validation} \label{sec:experimental_results}

Without loss of generality, we make use of a realistic example sufficiently complex for analyzing the ability to highlight functional inefficiencies of the proposed model that are made evindet by cold start events. It consists of online e-commerce application that we modeled according to the typical architecture of the AWS Lambda environment \cite{AWSLambda}.
Such applications are broadly structured as follows:

\begin{itemize}
    \item Handler Function: This function encodes the business logic. It is automatically invoked by the runtime of the deployment environment.
    \item Initialization functions: These functions are required for code started in cold mode. They are executed by the handler for each new instance (cold start). These can be used for initializing database connections, instantiate new objects, and load configuration files.
    \item Core Business Functions: These functions characterize the implemented FaaS architecture.
\end{itemize}

Table \ref{tab:business-functions} reports the list of functions that we considered, along with their logical aggregation used to define sagas. 
These functional entities allows defining topological cells, which are associated with $face$ of the graph. 

\begin{figure}[h!] 
    \centering
    \includegraphics[width=0.5\textwidth]{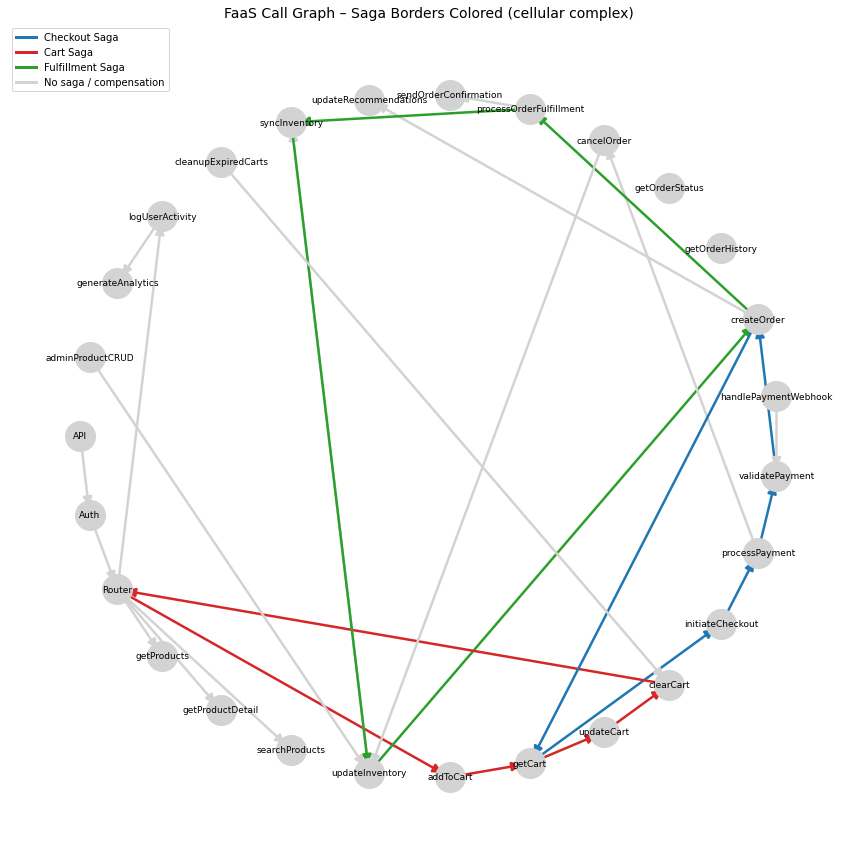} 
    \caption{Graph of the running example, including sagas.}
    \label{grafo_servizio} 
\end{figure}

In addition to the function aggregations listed in Table \ref{tab:business-functions}, in the model we introduced a compensation loop that is not handled like a saga. Beyond that, in the resulting service graph, shown in Figure \ref{grafo_servizio}, two functions are deliberately left isolated. The reason of this setting is that these functions are not immediately involved in the execution of sagas for interacting with the requesting clients and are called at a later time after service provision. This way, we can demonstrate the detection of isolated components in the service graph. These isolated functions are highlighted by null eigenvectors of $L_0=B_1B_1^T$ and Betti numbers, they increase the kernel size of $L_0$ and are not involved in the flows components generated by the Hodge decomposition.  
Figure \ref{grafo_servizio} highlights the faces associated with the defined sagas. It is worth to observe that when such such a service is deployed and orchestrated, a larger number of sagas is typically present. However, the combinatorial nature of the proposed approach is not impacted by the size of the graph and the used example allows demonstrating the detection capabilities of the proposal.
Through the proposed approach it is possible to analyze all the issues that  
are reported in Table \ref{tab:topological-cycles}. For this purpose it suffices to define co-chains over the function invocation graphs depending on the metrics reported in the table. Each row of Table \ref{tab:topological-cycles} reports the topological characterization of a workflow in the service graph, its topological model, the nature of the observable harmonic components along with the interested Betti number, and the affected observable metrics. In particular, $\beta_0 =1$ means causal/temporal cycle (1D) and only one component connected. It means that only one graph component is involved and when the action fails it can be re-executed in the same way. $\beta_1 =1$ indicates a true cycle, which could be traveled to infinity. Such $L_1$ harmonic captures its own energy trapped in the cycle. $\beta_2 =1$ means a topological 2D closed surface saga. It indicates that graph does not only have cycles, but also faces, that are 2D structure. 

The presence of a Harmonic component on $L_0$ indicates that nodes/states that cannot be eliminated since it is a static problem, in the location where energy is trapped. A Harmonic component on $L_1$ refers to cycles. The related problem is dynamic, it indicates an energy component that turns. If the Harmonic component is on $L_2$ it indicates a multidimensional problem, with potential multiple paths.

\begin{table}[t]
\centering
\caption{Examples of Core Business Functions for a Commercial Application}
\label{tab:business-functions}
\footnotesize
\begin{tabularx}{\columnwidth}{p{1.8cm}lp{7cm}}
\toprule
\textbf{Layer/Category} & \textbf{Functions} \\
\midrule
\textbf{API Layer} & 
API Gateway + Lambda \\
\midrule
\multirow{3}{1.8cm}{\textbf{Core Infrastructure Functions}} & 
Authentication/Authorization (1--2 functions) \\\\
& Routing/Orchestration (1 function) \\
\midrule
\multirow{13}{1.8cm}{\textbf{Core Business Functions}} & 
\textbf{Product Catalog:} \\
& \quad getProducts -- Product list \\
& \quad getProductDetail -- Product detail \\
& \quad searchProducts -- Search \\
& \quad updateInventory -- Inventory Management \\
& \\
& \textbf{Shopping Cart:} \\
& \quad addToCart -- Add to cart \\
& \quad getCart -- View cart \\
& \quad updateCart -- Modify cart \\
& \quad clearCart -- Clear cart \\
& \\
& \textbf{Checkout \& Payments:} \\
& \quad initiateCheckout -- Start checkout \\
& \quad processPayment -- Process Payment \\
& \quad validatePayment -- Validate payment \\
& \quad handlePaymentWebhook -- Webhook payments \\
& \\
& \textbf{Orders:} \\
& \quad createOrder -- Create order \\
& \quad getOrderHistory -- Order history \\
& \quad getOrderStatus -- Order Status \\
& \quad cancelOrder -- Cancel order \\
\midrule
\multirow{8}{1.8cm}{\textbf{Background Functions}} & 
\textbf{Event Processing (SQS/SNS/EventBridge):} \\
& \quad processOrderFulfillment -- Order fulfillment \\
& \quad sendOrderConfirmation -- Email confirmation \\
& \quad updateRecommendations -- Update recommendations \\
& \quad syncInventory -- Sync Inventory \\
& \\
& \textbf{Cron Jobs (EventBridge Scheduler):} \\
& \quad generateDailyReports -- Daily reports \\
& \quad cleanupExpiredCarts -- Cart cleaning \\
& \quad backupDatabase -- Data backup \\
\midrule
\multirow{3}{1.8cm}{\textbf{Support Functions}} & 
\textbf{Analytics:} \\
& \quad logUserActivity -- Activity tracking \\
& \quad generateAnalytics -- Analytics \\
& \quad abTestHandler -- A/B testing management \\
\midrule
\multirow{3}{1.8cm}{\textbf{System Functions}} & 
\textbf{Administration:} \\
& \quad adminProductCRUD -- CRUD Products (admin) \\
& \quad viewSystemMetrics -- System metrics \\
& \quad processBulkUpload -- Bulk upload \\
\bottomrule
\end{tabularx}
\end{table}

To evaluate the proposal it is necessary to consider the
the level of redundancy of the most popular deployment technologies, such as OpenFaaS over Kubernetes \cite{openfaasHome}. Considering that typically several instances of the same function are needed and that recycling them to create different sagas is considered good practice, the logical relationships generated by their interactions could become a tangle that is difficult to control.
In this situation, the analysis of events that can affect observable performance, such as cold starts, is of greatest importance. In our analysis we introduce cold starts in the flow definition $f$ and the optimal $M_1$ becomes a measure of how much control it takes to get closer to the target, which is the minimization of the Harmonic component due to cold start events.

The most direct way to stimulate the convergence of the proposed algorithm and observe its capabilities is to insulate and highlight a significant energy component in the edges of the compensation cycle \(("processPayment" \to "cancelOrder"),\) \(("cancelOrder" \to "updateInventory"),\) and \(("updateInventory" \to "syncInventory")\) is to make cold start happen in the involved functions.

For this analysis, we used co-chains that depend on both the functional load and the occurrence of cold start events. A co-chain \textit{f} is associated with the offered functional load. It represents a non-conservative flow of requests on the edges of the FaaS invocation graph. Each component of \textit{f} represents the average number of calls on a specific edge. Its values were initially randomly generated according to a Poisson distribution with mean 10. We also considered special edges with increased values. In particular, the edge $API\rightarrow Auth$ received an increment of 30 requests, simulating heavy traffic ingress from the front-end. The edge $processPayment \rightarrow validatePayment$ receives an increment of 15 requests, simulating flow amplification in some critical functions. The resulting \textit{f} is therefone non-conservative, as it could happens in real FaaS systems.
For what concerns the cold start events, the basic idea is to model the cold start using a weighted 1-cochain that measures induced latency. We therefore introduced a 1-cochain flow $f^{cs}$ whose value increases as the additional average latency on the edges increases due to cold start. More specifically, if the target function is in cold start then all incoming edges receive a high weight, while if it is warm the relative weight is 0. The co-chain combination by means of the point product results:
\begin{equation}
    f^{lat} = f^{req} \odot f^{cs}
\end{equation}
For our analysis $f^{cs}$ equals 30, 20, and 40 in all edges ending to the functions $processPayment$, $validatePayment$, and $syncInventory$, respectively.

The following Hodge decomposition can be interpreted as:
\begin{itemize}

    \item Gradient component:  "structural" cold start, with latencies due to functional hotspots, i.e. functions that receive many calls from different points and if a function is often cold and very central then it can generate a dominant gradient. These are functions to keep active at all times.

    \item Curl component: "orchestrated" cold start, with latencies circulating within a well-defined and managed saga. It means that the cold start is "trapped" inside the saga and it does not depend on external inputs, but on the cycle itself. This is a quantitative measure of how fragile a saga is at cold start and the relevant functions should be pre-warmed.

    \item Harmonic component: topological cold start that can be interpreted with latencies that do not create an accumulation of requests, are not explainable by any saga and that are typically due to criticalities present in compensation cycles, asynchronous webhooks, retry paths or operational loops outside the sagas. Systemic problems that need corrective action.
\end{itemize}

We implemented Algorithm \ref{alg:complete-recursive-iteration} with the following parameter values, decided by repeated experiments:
$\alpha = 10^{-3}$,
$M_{ref} = diag(1.0)$,
$M_{min} = 10^{-3}$,
$\lambda = 0.01$, 
$\beta = 1$.

\begin{figure}[h!] 
    \centering
    \includegraphics[width=0.5\textwidth]{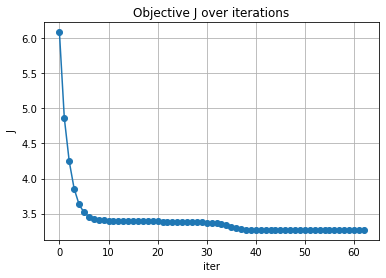} 
    \caption{Functional J value over iteration cycles.}
    \label{metrc_J} 
\end{figure}

Figure \ref{metrc_J} shows temporal evolution of the $J$ metric when the proposed procedure is applied. A significant convergence speed is observed, which indicates good ability to address situations characterized by high variability in the frequency of service requests.

\begin{figure}[h!] 
    \centering
    \includegraphics[width=0.5\textwidth]{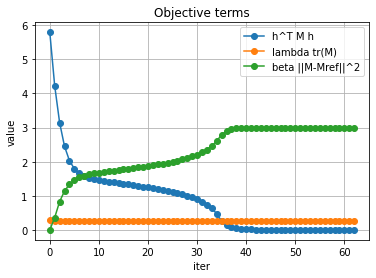} 
    \caption{Metric terms.}
    \label{metrc_terms} 
\end{figure}

It is also interesting to observe the effect of the optimization on the components that generate the metric, shown in Figure \ref{metrc_terms}. In particular, the evolution of the relative value between the three components shows the adaptation of the matrix $M_1$ on the three components of the metric and that the adaptation operates mainly on the harmonic component.

\begin{figure}[h!] 
    \centering
    \includegraphics[width=0.5\textwidth]{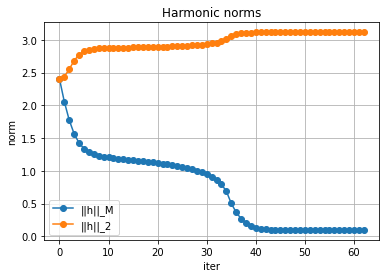} 
    \caption{Harmonic norms.}
    \label{harmonic_norms} 
\end{figure}

Focusing on the harmonic component, its role as a performance metric clearly emerges by observing Figure \ref{harmonic_norms}. Indeed, comparing the initial and final of harmonic norms induced by the initial and final $M_1$ matrices, it is evident that there has been a progressive filtering of the truly harmonic energy affecting the service, which is concentrated at the edges of the uncontrolled cycle. In other words, we are converging towards a representation free from numerical effects that could affect the identification of the Hodge flow components. 

\begin{figure*}[h!] %
    \centering
    \includegraphics[width=0.8\textwidth]{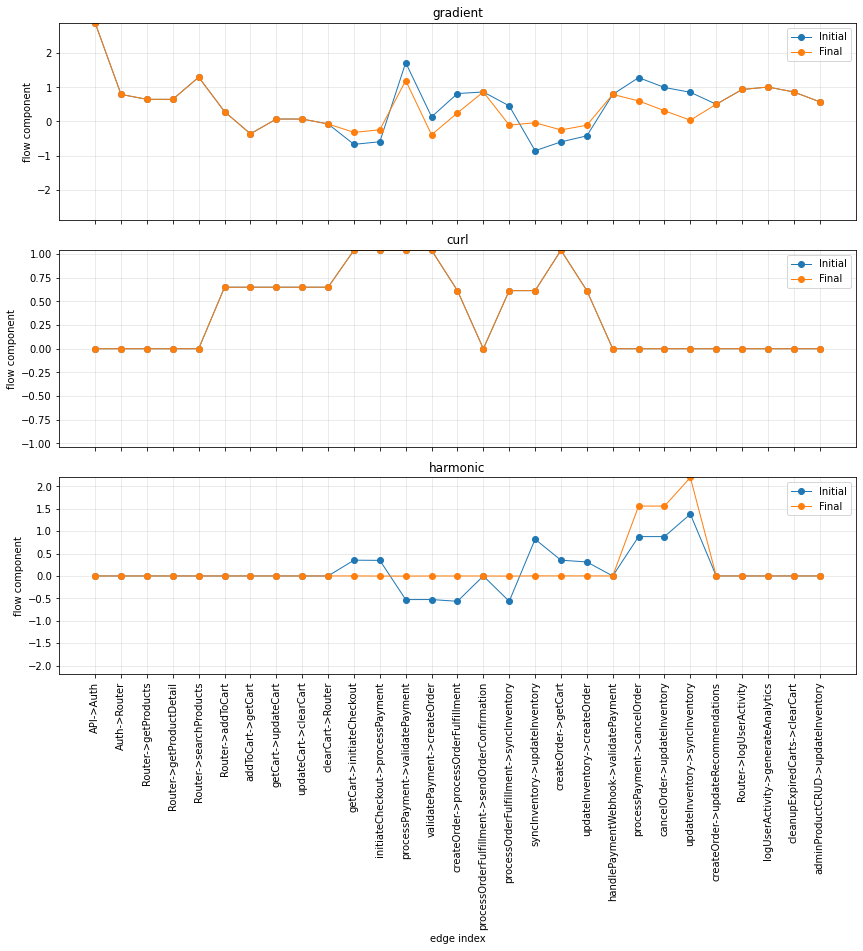} 
    \caption{Harmonic components.}
    \label{harmonic_processed} 
\end{figure*}

The final effect of the iteration process is evident in Figure \ref{harmonic_processed}. On some edges, what previously appeared as harmonic residues became explainable by other components (gradient) in the new metric. The observed redistribution of the harmonic component toward the compensation and inventory synchronization edges, together with the stability of the curl component associated with well-defined sagas, is fully consistent with the expected structural behavior of the system under the introduced cold-start perturbation. 

\begin{figure}[h!] 
    \centering
    \includegraphics[width=0.5\textwidth]{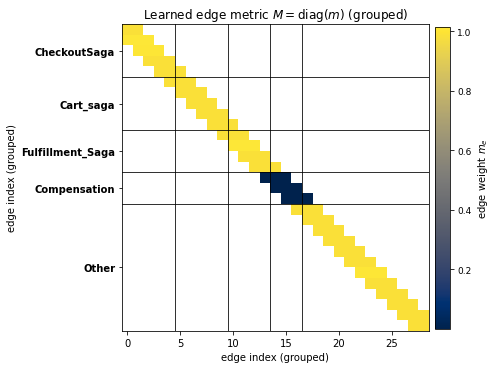} 
    \caption{Learned metric.}
    \label{metric} 
\end{figure}

The action of the learned metric over workflows can be better appreciated in Figure \ref{metric}. The compensation saga has been properly identified and valued. Thus, the flow that passes through the compensation cycle can be coherently explained as part of a saga (curl) or as a gradient of a potential, once the metric reflects the real workings.

\section{Conclusion} \label{sec:conclusion}

This paper presents a procedure for identifying metric deviations in serverless platforms by analyzing the information flows between functions. We employ Hodge decomposition to separate these operational flows into locally correctable components and globally persistent harmonic modes.
In particular, the harmonic modes arise from the inherent difficulty in precisely controlling function placement and fine-grained concurrency, causing the system to react through threshold-based mechanisms.
To address these deviations, our method performs a comprehensive analysis of the inter-function information flows. 
In order to correctly decompose flows, we propose an iterative method aiming to introducing a suitable metric, through which the harmonic component vanishes on many edges of the main sagas. This indicates that those paths are not structurally problematic and the traffic crossing them is explainable and manageable. The residual harmonic component instead concentrates on extra-saga loops, which represent true points of architectural fragility.

Future work will incorporate additional topological models to analyze harmonic components that emerge from resource contention among diverse services within a shared data center.

\bibliographystyle{IEEEtran}
\bibliography{references}

\end{document}